\documentstyle[multicol,aps,prb,epsf]{revtex}
\newcommand{\Vec}[1]{\mbox{\boldmath$#1$}}
\begin{document}

\draft

\title{
Magnetic-field induced triplet superconductivity 
in the Hubbard model on a triangular lattice
}

\author{
Ryotaro Arita, Kazuhiko Kuroki$^1$, and Hideo Aoki
}

\address{Department of Physics, University of Tokyo, Hongo,
Tokyo 113-0033, Japan}
\address{$^1$Department of Applied Physics and Chemistry,
University of Electro-Communications, Chofu, Tokyo 182-8585, Japan}

\date{\today}

\maketitle

\begin{abstract}
We propose theoretically that a magnetic field can realize 
spin-triplet superconductivity in repulsively interacting 
electron systems having strong ferromagnetic spin fluctuations.  
We confirm the general idea 
for the low-density Hubbard model on a triangular lattice,  
whose Fermi surface consists of disconnected pieces, 
by calculating the pairing susceptibility in a moderate magnetic field 
with the quantum Monte-Carlo method combined with the dynamical 
cluster approximation.   
\end{abstract}

\medskip

\pacs{PACS numbers: 71.10Fd, 74.25.Dw, 74.25.Ha}

\begin{multicols}{2}
\narrowtext
Triplet superconductivity is receiving current interests as 
one of the most fascinating new frontiers in superconductivity.  
Although there are in fact mounting experimental evidences for 
triplet pairing in various materials including 
an organic conductor TMTSF\cite{TMTSF1,TMTSF2},
heavy fermion systems\cite{UPt3}, or a ruthenate\cite{SRO1,SRO2},   
theoretical understanding is far from straightforward.  
Among a variety of possible approaches to realize triplet pairing,
the mechanism which exploits the exchange of ferromagnetic 
spin fluctuation has long been studied, dating back to
the paramagnon interpretation of superfluid $^3$He 
in the 1960's\cite{Anderson,Appel}.  Recent discoveries 
of UGe$_2$\cite{UGe2}, ZrZn$_2$\cite{ZrZn2} or UGeRh\cite{UGeRh} 
where the superconductivity {\it coexists} with ferromagnetism 
have invoked a renewed interest.

One big question about the triplet pairing is how 
to raise its $T_C$.  
While singlet superconductivity mediated by antiferromagnetic 
spin fluctuation almost certainly enjoys its realization
in high-$T_C$ cuprates, 
triplet superconductivity mediated by ferromagnetic spin fluctuation 
should in general have lower $T_c$'s. 
This is mainly because spin a triplet pair can only exploit 
one component (e.g. longitudinal [$zz$] one for an 
$S^z=1$ pair) in the spin fluctuation 
as the pairing interaction, 
while a singlet pair can exploit all the three 
(two transverse [$+-$] as well as longitudinal) 
components\cite{Arita,Monthoux}. 

To overcome this difficulty, various authors have proposed various ideas. 
Monthoux {\it et al.}\cite{Monthoux2} have shown phenomenologically that 
the Ising-like anisotropy in (ferromagnetic) spin fluctuations favors 
the triplet superconductivity in the $zz$ channel\cite{Ogata}. 
Here we can notice that the anisotropy can be readily realized 
if we introduce an 
external magnetic field. Namely, when the ferromagnetic spin fluctuation 
is dominant, the spin susceptibility 
$\chi_{zz} = \int d\tau \langle S^z(\tau)S^z(0) \rangle$ 
is enhanced and an Ising-like anisotropy will be realized in 
a moderate magnetic field $\Vec{B}$ (assumed here to be 
$\parallel \hat{z}$).  Thus here we test the idea 
by studying the effect of a magnetic field to the triplet pairing instability 
in the ferromagnetic microscopic models rather than a phenomenology.

On the other hand, motivated by the ferromagnetic superconductivity 
in UGe$_2$\cite{UGe2}, ZrZn$_2$\cite{ZrZn2} and UGeRh\cite{UGeRh}, 
Kirkpatrick {\it et al.}\cite{Kirkpatrick} have proposed that the 
ferromagnetic spin-fluctuation mediated superconductivity 
should have a higher $T_c$ in the ferromagnetic phase than in 
the paramagnetic phase. With field-theoretic techniques they 
have shown phenomenologically that the coupling of longitudinal 
spin fluctuations with transverse spin waves plays a crucial role
to raise $T_c$.

Thus coexisting ferromagnetism and superconductivity should also be an 
interesting avenue, but theoretical realization of this is again 
not straightforward. Namely, the present authors have systematically 
studied the Hubbard model on various two-(2D) and 
three-dimensional(3D) lattice structures 
with the fluctuation exchange (FLEX) approximation, and have 
concluded that 2D systems are much more favorable for superconductivity 
than 3D systems\cite{Arita}. 
Physical reason for this has been identified in a much larger 
volume fraction of the phase space over which the interaction 
contributes to the Eliashberg equation. 
A similar conclusion has been obtained independently by Monthoux 
{\it et al} from a phenomenological calculation\cite{Monthoux,Nakamura}.  
This should also apply to the ferromagnetic superconductivity, 
but a serious problem in 2D 
is the Curie temperature $T_c$, which would 
upper-bound the ferromagnetic superconductivity, 
should be much lower\cite{fnTc} in 2D 
due to quantum fluctuations.   

This is exactly why we consider applying a magnetic field to 
an ``almost ferromagnetic" 
2D system to stabilize the ferromagnetism. So, in sharp contrast to a 
conventional wisdom that superconductivity and magnetic field are inimical 
to each other, we are here talking about a unique situation where the pairing
arises {\it due to} the magnetic field\cite{UjiComm}. 
To repeat the idea, we consider a microscopic model close to ferromagnetism, 
which should become spin-polarized in a relatively small magnetic field even at 
high temperatures. Then we can possibly exploit the favorable modes {\it a la} 
Kirkpatrick {\it et al.} without discarding the advantage of 2D. 

\begin{figure}
\begin{center}
\leavevmode\epsfysize=55mm \epsfbox{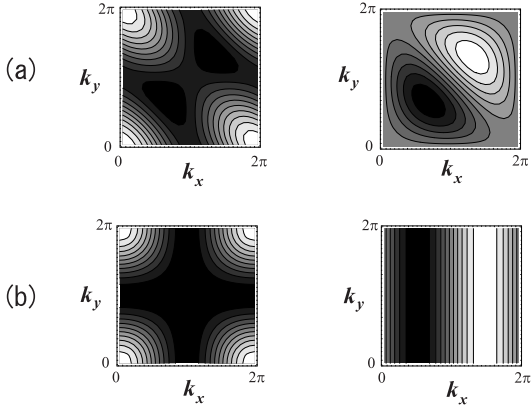}
\caption{Gray-scale plots for the band dispersion (left panels)
and the dominant gap function (right)
for the Hubbard model on the triangular lattice (a) and
$t$-$t'$ lattice (b). }
\label{FSGAP}
\end{center}
\end{figure}

As for the microscopic model we take here the repulsive Hubbard model.  
While it has been shown, as mentioned above, that triplet 
pairing is hard to realize in the Hubbard model in ordinary 
lattices\cite{Arita}, a notable 
wayout has been proposed by two of the present authors\cite{Kuroki}, 
where the triplet pairing is shown to become dominant if we 
consider systems having Fermi surfaces consisting of 
{\it disconnected pieces}.  There, we can insert the nodal 
line(s) that are required to exist in a triplet gap function, 
in between the pieces, which greatly helps in enhancing $T_C$.   
A simplest example where such a Fermi surface 
occurs is the double-dipped band dispersion in 
the triangular lattice (Fig.\ref{FSGAP}) when the 
electron concentration is dilute.  
Ref.\cite{Kuroki} has shown that 
an f-wave triplet pairing is indeed much favored than usual. 
However, it is still a subtle problem 
whether we can obtain a finite $T_c$.  
Thus it is a theoretical challenge to clarify whether the triplet 
superconductivity in the $zz$ channel (i.e., the spins of the 
pair $\parallel \Vec{B}$ with the d vector $\Vec{d} \perp \hat{z}$) 
can become dominant.  
If such a triplet instability is enhanced by 
applying an external magnetic field, it will open a new avenue 
for the triplet 
superconductivity\cite{UjiComm}.
We have calculated the triplet pairing susceptibility in magnetic 
fields, and show that the magnetic-field induced triplet superconductivity 
should exist, although we have not been able to quantitatively 
estimate $T_C$.  


From the motivation of the study, the method should 
be able to handle, in a microscopic way, 
higher-order processes such as 
the coupling between longitudinal spin fluctuations with 
transverse spin waves, and pairing fluctuations in low dimensions. 
Here we take the quantum Monte-Carlo (QMC) method in combination with 
the dynamical cluster approximation (DCA) formulated recently by 
Jarrell {\it et al}.\cite{Jarrell1,Jarrell2,Jarrell3}. 
In DCA, the Hubbard model is mapped to a self-consistently embedded
cluster of Anderson impurities rather than a single impurity considered in
the dynamical mean field approximation (DMFA)\cite{DMFAreview}, so that 
DCA systematically incorporates nonlocal spatial fluctuations.  
Specifically, DCA+QMC, being a non-perturbative approach, 
automatically includes the higher order processes.  
We have solved the the cluster problem generated by the DCA using
the QMC method\cite{SuzukiTrotter} with the 
algorithm proposed by Hirsch and Fye\cite{Hirsch}. 
We choose the cluster size $N_c=4\times 4$, which is large 
enough to treat the anisotropic superconductivity\cite{supcomm}.
As for the effect of the magnetic field, we concentrate here on the 
Zeeman effect by applying the magnetic field $\Vec{B}$ parallel to the 
plane to ignore the orbital effect. 

The Hubbard Hamiltonian is 
\[
{\cal H} = -\sum_{ij\sigma}t_{ij} c^\dagger_{i\sigma}c_{j\sigma} 
+U\sum_{i}n_{i\uparrow}n_{i\downarrow} + h\sum_{i\sigma} {\rm sgn}(\sigma) 
c_{i\sigma}^{\dagger}c_{i\sigma}
\]
in the standard notation, where $t_{ij}$ is nonzero only 
between nearest neighbors, $h \equiv g\mu_B B$
is the Zeeman energy and the spin quantization axis 
taken to be $\parallel \Vec{B}$. Hereafter, the Coulomb 
repulsion $U$ is set equal to the band width $W$, 
which is taken to be 2 here.  

The one-electron band dispersion, 
$\varepsilon=2t[\cos(k_x)+\cos(k_x)+\cos(k_x+k_y)]$, for
the triangular lattice (Fig. \ref{FSGAP}) has two pockets 
for the Fermi surface 
when the band is less than quarter-filled ($n<0.5$).  
For comparison with the case where ferromagnetic fluctuations 
dominate but the Fermi surface is simply connected, 
we also study the low-density Hubbard model on a square 
lattice with next-nearest transfers ($t$-$t'$ lattice hereafter).
Its dispersion is 
$\varepsilon=2t[\cos(k_x)+\cos(k_x)]+2t'\cos(k_x)\cos(k_y)$, 
for which we take $t'=0.5|t|$ to make the the density of states 
diverge at the band bottom to realize strong ferromagnetic 
fluctuations\cite{Arita,Hlubina,Arita2}. 
The Fermi surface for this lattice is a simply-connected star shape 
when dilute. To facilitate comparison with triangular 
lattice we adjust $t$ to make $W=2 (=U)$.

\begin{figure}
\begin{center}
\leavevmode\epsfysize=40mm \epsfbox{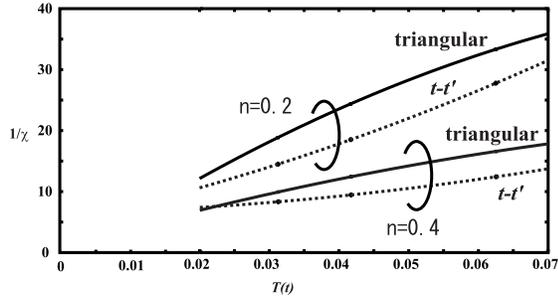}
\caption{The inverse of uniform spin susceptibilities versus temperature 
for the triangular and $t$-$t'$ lattices with $n=0.2, 0.4$.}
\label{spinsus02}
\end{center}
\end{figure}

Let us now present the results. We start by confirming 
that the systems considered here all have strong
ferromagnetic fluctuations. In Fig. \ref{spinsus02}, 
we plot the inverse spin susceptibility($\chi$) versus temperature 
for the triangular and $t$-$t'$ lattices. 
We can see that both systems approach ferromagnetism 
(i.e., $\chi$ increases) for $T\rightarrow 0$.  
The tendency is stronger for a dilute filling ($n=0.2$) with a disconnected 
Fermi surface than for a near-quarter filling ($n=0.4$) for which 
the Fermi surface is almost simply-connected.  

\begin{figure}
\begin{center}
\leavevmode\epsfysize=40mm \epsfbox{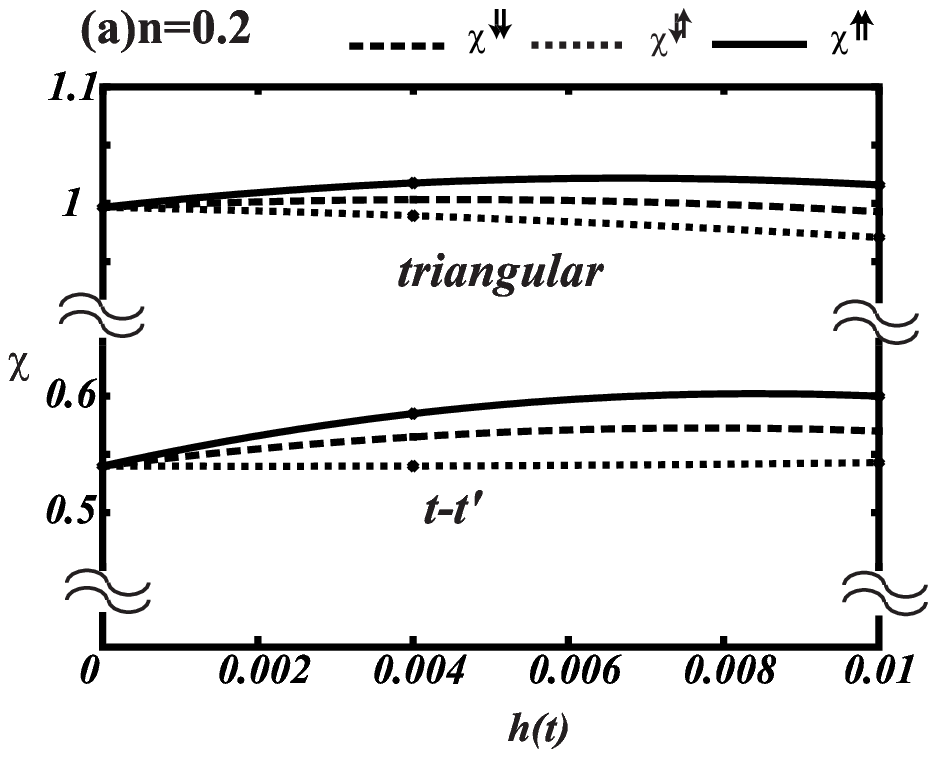}
\leavevmode\epsfysize=40mm \epsfbox{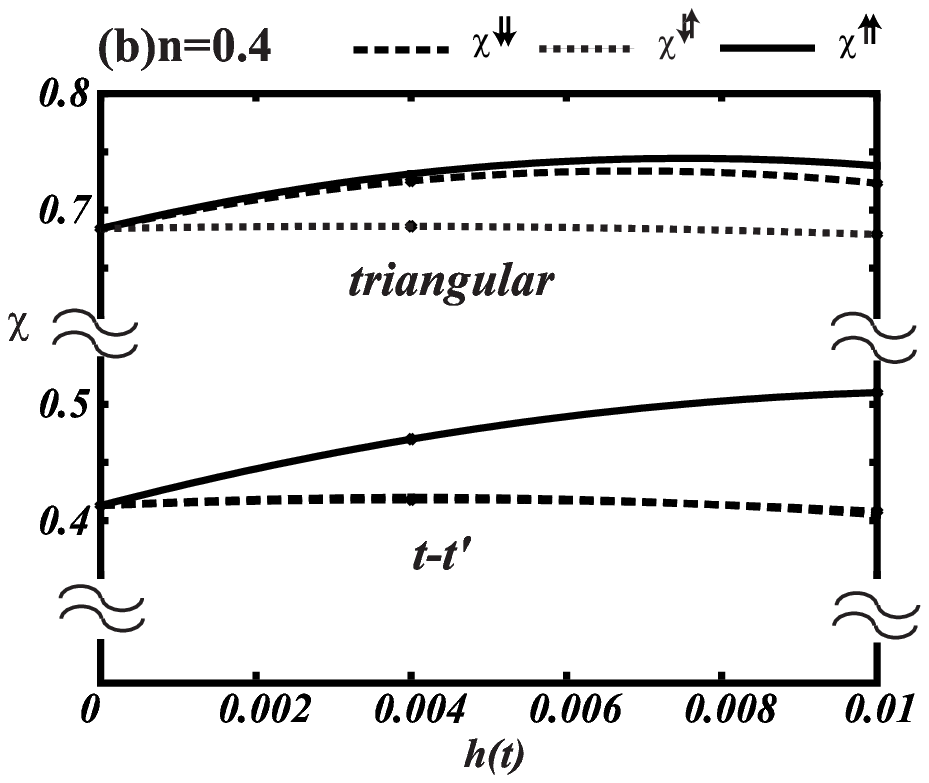}
\caption{The triplet pairing susceptibilities versus magnetic field 
for the triangular and $t$-$t'$ lattices for $n=0.2$(a)
and $n=0.4$(b).}
\label{pairmag02}
\end{center}
\end{figure}

In the presence of the Zeeman energy the triplet pairing susceptibilities, 
$\chi^{\uparrow\uparrow} \sim \langle 
c^\dagger_\uparrow c^\dagger_\uparrow c_\uparrow c_\uparrow
\rangle$, 
$\chi^{\uparrow\downarrow} \sim \langle
c^\dagger_\uparrow c^\dagger_\downarrow c_\downarrow c_\uparrow
\rangle$, 
$\chi^{\downarrow\downarrow} \sim \langle
c^\dagger_\downarrow c^\dagger_\downarrow c_\downarrow c_\downarrow
\rangle$ with $S_z^{\rm total}=1,0,-1$ respectively, differ from one another.  
Figure \ref{pairmag02} plots these versus external magnetic field 
at temperature $T(\equiv 1/\beta)=1/32$.
For the two Fermi pockets in the 
triangular lattice we can consider an $f$-wave (with three nodes) 
gap function, 
$\phi=\sqrt{3/32}
[\cos(k_x/2)\cos(k_y/2)\sin((k_x-k_y)/2)]$.  
For this pairing symmetry all the nodes avoid intersecting the Fermi surface, 
so that the pair-scattering processes all around the Fermi surface(FS)
has nonzero contributions to the pairing vertex, 
\[
-\frac{\sum_{\Vec{k},\Vec{k}' \in {\rm FS}}V(\Vec{k}-\Vec{k}')
\phi(\Vec{k})\phi(\Vec{k}')}
{\sum_{\Vec{k} \in {\rm FS}}\phi^2(\Vec{k})},  
\]
which measures the tendency toward the pairing instability. 
We can see that the $f$-wave susceptibilities of
triangular lattice for $n=0.2$ are indeed much larger than
those for the $t$-$t'$ lattice [with $\phi=\sqrt{1/2}\sin(k_x)$].  
$\chi$ is larger for the dilute band ($n=0.2$).   
These are consistent with the results of our previous
calculation by FLEX\cite{Kuroki}.

As for the magnetic field dependence, we can see 
that $\chi^{\uparrow\uparrow}$ is indeed enhanced in the presence 
of moderate magnetic fields in all the cases studied.  
The enhancement, however saturates for a larger magnetic field 
($h\sim 1.0 \times 10^{-2}$), so that the enhancement 
is peaked at some $h (\equiv h_{\rm max}$). This is also the case with the most 
promising candidate, the low-density triangular lattice. 
This can be interpreted as follows. While the magnetic field does 
contribute to $\chi^{\uparrow\uparrow}$ from both the Kirkpatrick processes 
and the anisotropy in the spin fluctuations, the Fermi surface for 
the majority spin expands and becomes simply-connected when the 
Zeeman energy is too large. The nodes in $\phi$ will then have to intersect 
the Fermi surface, which should degrade the pairings of majority spins.
Therefore, the magnetic-field induced superconductivity can be
expected only in the appropriate magnetic field.

\begin{figure}
\begin{center}
\leavevmode\epsfysize=40mm \epsfbox{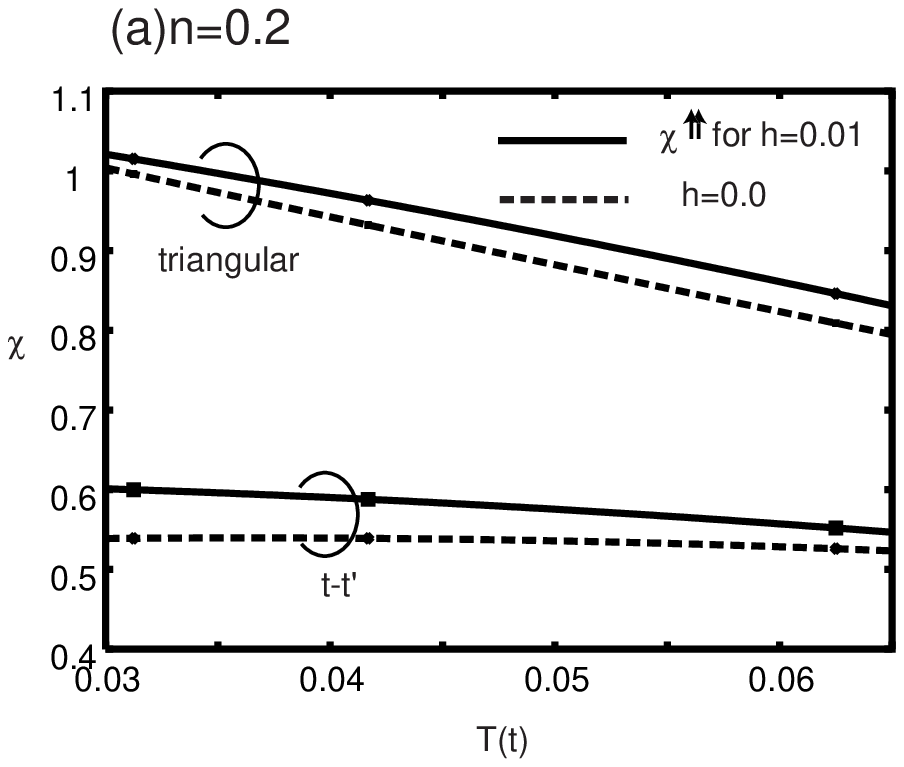}
\leavevmode\epsfysize=40mm \epsfbox{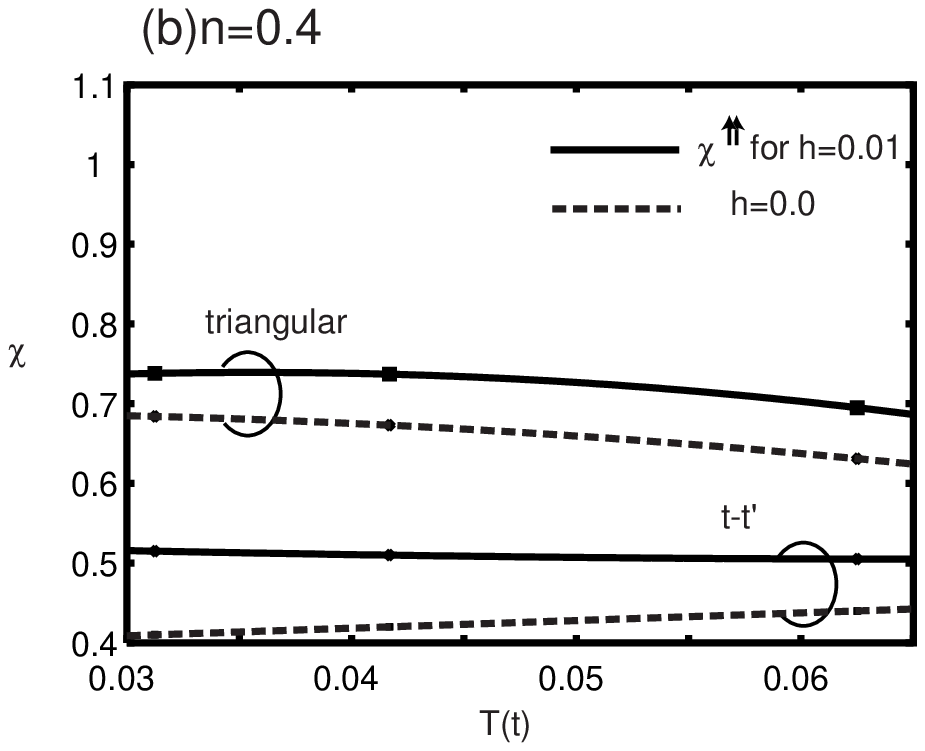}
\caption{The triplet pairing susceptibilities versus temperature 
for the triangular and $t$-$t'$ lattices for $n=0.2$(a) 
and $n=0.4$(b).}
\label{pairtem02}
\end{center}
\end{figure}

The ``appropriate" magnetic field is dictated by the temperature.  
So we move on to the $T$-dependence of the pairing susceptibilities. 
We focus on $\chi^{\uparrow\uparrow}(h=1.0 \times 10^{-2})$ and 
$\chi(h=0)$ in Fig. \ref{pairtem02}.  
$\chi(h=0)$ grows toward $T=0$ for the triangular lattice with $n=0.2$, 
which is again consistent with the FLEX calculation\cite{Kuroki}. 
As for the enhancement of $\chi^{\uparrow\uparrow}(h=1.0 \times 10^{-2})$ 
above $\chi(h=0)$, the enhancement decreases  toward $T=0$ for the dilute 
triangular lattice.  
Extending the above argument we can see that this should be the case. 
For a {\it fixed} magnetic field the spin polarization becomes larger 
as the temperature is decreased, so that the majority-spin Fermi surface again 
becomes simply-connected. This implies that 
$\chi^{\uparrow\uparrow}$ continues to be enhanced 
at low temperatures only when 
we decrease the magnetic field accordingly.  

\begin{figure}
\begin{center}
\leavevmode\epsfysize=40mm \epsfbox{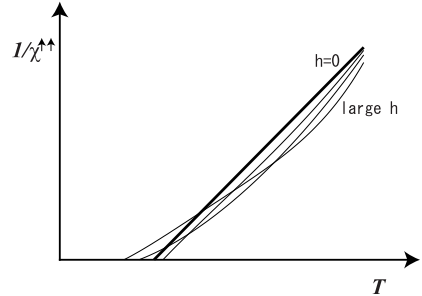}
\caption{A schematic $1/\chi^{\uparrow\uparrow}$ versus $T$ 
for various values of magnetic field.  
The bold line represents $h=0$.}
\label{schematic}
\end{center}
\end{figure}

The situation is summarized in Fig. \ref{schematic}, 
which schematically depicts the temperature-dependence 
of $1/\chi^{\uparrow\uparrow}$.  
A bunch of lines represent the behavior for various values 
of $h$, where $\chi (1/\chi)$ at each $T$ 
first increases (decreases) 
and then reversed when $h$ exceeds $h_{\rm max}$, 
where $h_{\rm max}$ decreases with $T$.   
If we assume that the derivative of $\chi$ with respect to $h$ 
continues to be positive around $T_C$ at which $\chi(h=0)$ diverges, 
the bunch of lines forms an envelope in such a way that 
the $T_C$ is enhanced 
for an intermediate value ($h\sim 0.01$) of the magnetic field.  
The value $h=0.01$ corresponds to $O(10T)$ 
for the band width $\sim O(1$ eV). 

From the above argument a system in which the Fermi surface remains 
disconnected for larger Zeeman splitting may seem more favorable.  
However, we have to have an almost flat bottomed dispersion 
to realize ferromagnetism to begin with, so the persistent disconnectivity 
might be incompatible with this.  Whether this can be overcome is an 
interesting future work. 
In conclusion, we have suggested an exotic possibility of 
a magnetic-field induced triplet superconductivity.
This work was supported in part by a Grant-in-aid for scientific 
research and Special coordination funds from the ministry of 
education of Japan. Numerical calculations were performed on 
SR8000 in ISSP, University of Tokyo.

\end{multicols}

\begin{references}
\bibitem{TMTSF1}I.J. Lee, A.P. Hope, M.J. Leone, and M.J. Naughton,
Synth. Met. {\bf 70}, 747 (1995).
\bibitem{TMTSF2}I.J. Lee, M.J. Naughton, G.M. Danner and P.M. Chaikin, 
Phys. Rev. Lett. {\bf 78}, 3555 (1997); 
I.J. Lee, P.M. Chaikin and M.J. Naughton, Phys. Rev. B {\bf 62}, R14669 (2000).
\bibitem{UPt3}H. Tou, Y. Kitaoka, K. Asayama, H. Kimura, Y. Onuki,
E. Yamamoto, and K. Maezawa, Phys. Rev. Lett. {\bf 77}, 1374 (1996).
\bibitem{SRO1}G.M. Luke, Y. Fudamoto, K.M. Kojima, M.I. Larkin, 
J. Merrin, B. Nachumi, Y.J. Uemura, Y. Maeno,  Z.Q. Mao, Y. Mori, 
H. Nakamura, and M. Sigrist, Nature {\bf 394}, 558 (1998).
\bibitem{SRO2}K. Ishida, H. Mukuda, Y. Kitaoka, K. Asayama, Z.Q. Mao,
Y. Mori, and Y. Maeno, Nature {\bf 396}, 658 (1998).
\bibitem{Anderson}K.A. Brueckner, T. Soda, P.W. Anderson and P. Morel,
Phys. Rev. {\bf 118}, 1442 (1960); P.W. Anderson and W.F. Brinkman, 
Phys. Rev. Lett. {\bf 30}, 1108 (1973).
\bibitem{Appel}D. Fay and J. Appel, Phys. Rev. B {\bf 22}, 3173(1980), 
and references therein.
\bibitem{UGe2}S.S. Saxena, P. Agarwal, K. Ahilan, F.M. Grosche,
R.K.W. Haselwimmer, M.J. Steiner, E. Pugh, I.R. Walker, S.R. Julian,
P. Monthoux, G.G. Lonzarich, A. Huxley, I. Sheikin, D. Braithwaite, and 
J. Flouquet, Nature {\bf 406}, 587 (2000).
\bibitem{ZrZn2}C. Pfleiderer, M. Uhlarz, S.M. Hayden, R. Vollmer, 
H. von Lohneysen, N.R. Bernhoeft, and G.G. Lonzarich, 
Nature {\bf 412}, 58 (2001).
\bibitem{UGeRh}D. Aoki, A. Huxley, E. Ressouche, D. Braithwaite, 
J. Flouquet, J.P. Brison, E. Lhotel, and C. Paulsen,
Nature {\bf 413}, 613 (2001).
\bibitem{Arita} R. Arita, K. Kuroki, and H. Aoki, Phys. Rev. B {\bf 60}, 
14585 (1999); J. Phys. Soc. Jpn. {\bf 69}, 1181 (2000).
\bibitem{Monthoux} P. Monthoux and G.G. Lonzarich, Phys. Rev. B
{\bf 63}, 054529 (2001). 
\bibitem{Monthoux2} P. Monthoux and G.G. Lonzarich, Phys. Rev. B
{\bf 59}, 14598 (1999). 
\bibitem{Ogata} As for the magnetic anisotropy in the spin 
fluctuations having finite wave numbers, T. Kuwabara and M. Ogata
[Phys. Rev. Lett. {\bf 85}, 4586 (2000)] and M. Sato and M. Kohmoto
[J. Phys. Soc. Jpn. {\bf 69}, 3505 (2000)] have studied 
its effect in the context of the superconductivity observed 
in Sr$_2$RuO$_4$, and have shown that the magnetic anisotropy 
indeed favors triplet pairing in the $\Vec{d} \perp \hat{z} 
(S_z=0$) channel.
\bibitem{Kirkpatrick} T.R. Kirkpatrick, D. Belitz,
T. Vojta, and R. Narayanan, Phys. Rev. Lett. {\bf 87},
127003 (2001); T.R. Kirkpatrick and D. Belitz, cond-mat/0204440.
\bibitem{Nakamura} S. Nakamura, T. Moriya and K. Ueda
[J. Phys. Soc. Jpn. {\bf 65}, 4026 (1996)] 
have performed a phenomenological calculation 
to conclude that $T_c$ in 3D systems 
is similar to that of 2D systems. There, the frequency spread of 
spin fluctuations is assumed to scale with the band width, while 
the FLEX result indicates that the spread scales with $t$.
\bibitem{fnTc}It should be noted that while both magnetic and
pairing transition temperatures are zero to be exact in 2D systems, 
here we expect that the coherence length for the pairing is 
much larger than that of the magnetism.
\bibitem{UjiComm} There are a variety of exotic 
magnetic-field-induced organic superconductors: see 
J.A. Symington {\it et al}, J. Phys. Condens. Mat. {\bf 12}, L641 (2000); 
S. Uji {\it et al}, Nature {\bf 410}, 908 (2001); 
N. Harrison {\it et al}, J. Phys. Condens. Mat. {\bf 13}, L389 (2001).
\bibitem{Kuroki} K. Kuroki and R. Arita, Phys. Rev. B, {\bf 63},
174507 (2001).
\bibitem{Jarrell1}M.H. Hettler, A.N. Tahvildar-Zadeh, M. Jarrell, 
T. Pruschke, and H.R. Krishnamurthy, Phys. Rev. B, {\bf 58}, 7475 (1998).
\bibitem{Jarrell2}M.H. Hettler, M. Mukherjee, M. Jarrell and H.R.
Krishnamurthy, Phys. Rev. B, {\bf 61}, 12739 (2000).
\bibitem{Jarrell3}Th. Maier, M. Jarrell T. Pruschke, and J. Keller,
Euro. Phys. J. B {\bf 13}, 613 (2000).
\bibitem{DMFAreview}A. Georges, G. Kotliar, W. Krauth, and M.J. Rozenberg,
Rev. Mod. Phys. {\bf 68}, 13 (1996).
\bibitem{SuzukiTrotter} 
As for the Suzuki-Trotter decomposition number, we take 32.
\bibitem{Hirsch}J.E. Hirsch and R.M. Fye, Phys. Rev. Lett. {\bf 56},
2521 (1986).
\bibitem{supcomm} We have calculated the pairing susceptibility 
for the self-consistent cluster, which should 
coincide with that for the original lattice for 
$N_c \rightarrow \infty$.
\bibitem{Hlubina}R. Hlubina, S. Sorella, and F. Guinea,
Phys. Rev. Lett. {\bf 78}, 1343 (1997).
\bibitem{Arita2}R. Arita, K. Kuroki, and H. Aoki, Phys. Rev. B {\bf 61},
3207 (2000).
\end{references}
\end{document}